\documentclass[usenatbib]{mnras}
\usepackage{color}
\usepackage{graphicx,graphics}
\usepackage{amssymb,amsmath}
\usepackage{natbib}
\usepackage {threeparttable}

\newcommand{\Lya}{Ly$\alpha$ }
\newcommand{\ourq}{J2329$-$0301}

\title[The evolution of the CGM around QSOs]
 {
Possible evolution of the circum-galactic medium around QSOs with QSO age and cosmic time revealed by \Lya halos
}

\author[R. Momose et al.]{
Rieko~Momose$^{1,2}$,
Tomotsugu~Goto$^{2}$,
Yousuke~Utsumi$^{3}$,
Tetsuya~Hashimoto$^{2}$, 
\newauthor
Chia-Ying Chiang$^{2}$,
Seong-Jin Kim$^{2}$,
Nobunari~Kashikawa$^{1}$,
Kazuhiro~Shimasaku$^{1,4}$,
\newauthor
Satoshi~Miyazaki$^{5}$
\\
$^1$
Department of Astronomy, School of Science,
The University of Tokyo,
7-3-1 Hongo, Bunkyo-ku, Tokyo 113-0033, JAPAN \\
$^2$
Institute of Astronomy, National Tsing Hua University,
101, Section 2 Kuang-Fu Road, Hsinchu, Taiwan, 30013 \\
$^3$
Kavli Institute for Particle Astrophysics and Cosmology (KIPAC),
SLAC National Accelerator Laboratory, Stanford University, \\
2575 Sand Hill Road, Menlo Park, CA  94025, USA \\
$^4$
Research Center for the Early Universe, The University of Tokyo, 
7-3-1 Hongo, Bunkyo-ku, Tokyo 113-0033, Japan \\
$^5$
National Astronomical Observatory of Japan, 
2-21-1 Osawa, Mitaka, Tokyo 181-8588, Japan
}

\date{}

\pagerange{\pageref{firstpage}--\pageref{lastpage}} \pubyear{2018}

\begin{document}

\label{firstpage}

\maketitle

\begin{abstract}
{
We first present new Subaru narrow-band observations of the \Lya halo around the quasi-stellar object (QSO) CFHQ J232908$-$030158 at $z=6.42$, 
which appears the most luminous and extended halo at $z>5$ ($L_{Ly\alpha}=9.8\times10^{43}$ erg s$^{-1}$ within $37$ pkpc diameter).
Then, combining these measurements with available data in the literature, we find two different evolutions of QSOs' \Lya halos. 
First is a possible short-term evolution with QSO age seen in four $z>6$ QSOs. 
We find the anti-correlation between the \Lya halo scales with QSOs’ IR luminosity, with \ourq's halo being the brightest and largest.
It indicates that ionizing photons escape more easily out to circum-galactic regions when host galaxies are less dusty.
We also find a positive correlation between IR luminosity and black hole mass ($M_\text{BH}$).
Given $M_\text{BH}$ as an indicator of QSO age, we  propose a hypothesis that a large \Lya halo mainly exists around QSOs in the young phase of their activity due to a small amount of dust. 
The second is an evolution with cosmic time seen over $z\sim2-5$.
We find the increase of surface brightness toward lower-redshift with a similar growth rate to that of dark matter halos (DHs) which evolve to
$M_\text{DH}=10^{12}-10^{13}$ M$_\odot$ at $z=2$.
The extent of \Lya halos is also found to increase at a rate scaling with the virial radius of growing DHs,
$r_\text{vir} \propto M_\text{DH}^{1/3}$($1+z$)$^{-1}$.
These increases are consistent with a scenario that the CGM around QSOs evolves in mass and size keeping pace with hosting DHs.
}
\end{abstract}


\section{Introduction}
\label{introduction}

Gas exchanges between galaxies and the inter-galactic medium (IGM) around them play a central role in star formation and galaxy evolution.
While the star formation activity of galaxies is maintained by gas inflows from the IGM (e.g., \citealp{deke09a, deke09b}), outflows induced by star formation heat or even blow away the cold gas in galaxies, thereby suppressing subsequent star formation (e.g., \citealp{mori02, mori04, scan05, mori06, dave11}). Since inflow and outflow gas passing through or reaching circum-galactic regions makes up the circum-galactic medium (CGM) extending over up to a few hundred physical kpc (pkpc), physical properties of the CGM such as mass and spatial extent provide valuable insights into star formation and galaxy evolution.

To understand the physical properties of the CGM, absorption lines are often used as a tracer of the cold gas phase of the CGM (e.g., \citealp{henna06,prochas09,henna13,bouche13,bouche16,prochas17}). 
These studies have shown the presence of cold ($T\sim10^4$ K) and metal-enriched ($Z>0.1$ Z$_\odot$) gas extending to a few hundred pkpc (e.g., \citealp{henna13,procha13,procha14,farina13,farina14,john15,lau16}). Additionally, systemic surveys targeting pair QSOs (Quasars Probing Quasars) have also indicated optically thick \ion{H}{i} gas in the CGM (e.g., \citealp{procha14}). 
Although these previous studies have found several important properties of the CGM around galaxies, they have not provided full spatial distributions of the CGM because absorption lines usually provide parameters of one dimensional nature. 

Another tracer is the emission lines from the CGM, among which the strongest is the hydrogen Ly$\alpha$. 
If neutral hydrogen gas constituting the CGM gas is illuminated by ionizing or \Lya radiation produced in the hosting galaxy by star formation or AGN activities, it can be observed as a diffuse, extended \Lya halo (or called a \Lya nebula).
Good examples are seen in \Lya emitters (LAEs), galaxies with strong \Lya emission. LAEs are known to possess \Lya halos with scale lengths reaching $\sim10$ pkpc independent of redshift (for studies based on stacking analysis, see, e.g., \citealp{stei11, matsu12, feld13, momo14,momo16}; for individual-basis studies, see, e.g., \citealp{rauch08,wiso16,lecler17}).

Quasi-stellar objects (QSOs) have brighter \Lya halos than normal galaxies, making it possible to detect them on an individual basis especially at $z\sim2-3$ (e.g. \citealp{heckman91a, heckman91b, chris06, north12, henna13, roche14, borisova16b, fath16}).
Recently, the discovery of extremely giant \Lya halos over $>300$ pkpc has been reported (e.g., \citealp{cantalupo14, martin14, henna15, borisova16b,cai17,arrigoni18}).
\citet{cantalupo14} have detected a giant \Lya halo, named Slug Nebula, with an extent of $\sim460$ pkpc around a QSO at $z=2.28$. This halo extends beyond the virial radius of the associated dark matter halo, indicating that \Lya emission of such a giant nebula also traces the IGM gas. 
\citet{henna15} have also discovered a giant \Lya halo of $310$ pkpc extent covering four QSOs. These extremely extended \Lya halos indicate the presence of a widely spread CGM, and/or a large amount of cold gas around QSOs.
However, at least at $z\sim3$, QSOs' \Lya halos seem to have similar surface brightness (SB) profiles despite different luminosities and sizes. \citet{borisova16b} have found that the SB profile of the Slug Nebula is consistent with the mean SB profile of \Lya halos around $17$ $z\sim3$ QSOs.
Such similarity of SB profiles indicates that QSOs' \Lya halos have the same origin.

Observational studies of QSOs' \Lya halos have revealed another important trend, evolution with cosmic time, although based on samples over a relatively narrow redshift range of $z \sim 2-4$ and with different SB limits.
\citet{north12} have found that the SBs of \Lya halos at $z\sim4.5$ are $1-2$ order of magnitude fainter than those at $z\sim2-3$. 
\citet{ginolfi18} have found that halos at $z\sim5$ are smaller than at $z\sim3-4$.
Moreover, \citet{farina17} have suggested a decline in the total \Lya luminosity of halos with redshift over $2<z<7$ due to a change in gas properties and/or powering mechanisms of \Lya radiation. 
On the other hand, \citet{arrigoni18b} have found an opposite trend \citep{north12,farina17,ginolfi18} that the mean SB of radio-quiet QSOs increases with redshift from $z=2$ to $z=3$.
Furthermore, no evolution has been found in the \Lya halos luminosity over $2<z<4.5$ \citep{fath16}. These discrepancies may be attributed to different methodologies (e.g., SB limits, observation methods [spectroscopy or imaging], methods to measure sizes) in the literature.
Hence, in order to comprehensively understand the properties and evolution of QSO \Lya halos, an analysis with a large sample compiled under the same conditions is required.

In addition, the number of observations for $z>6$ QSO halos is still limited.
\citet{goto09} have for the first time reported a possible presence of a \Lya halo around a $z>6$ QSO, CFHQ J232908$-$030158 (hereafter \ourq) at $z=6.4$, by showing an extended feature in a $z'$-band image whose bandpass covers Ly$\alpha$, while detecting no such feature in other bands ($i'$ and $z_r$-bands).
After subtracting the PSF component from the QSO, they have confirmed residual emission of $4\arcsec$-wide corresponding to $\sim22$ pkpc around the QSO.
On the other hand, \citet{deca12} have not detected an extended \Lya halo around two QSOs at $z=6.3$ and $6.4$ in spite of using deep Hubble Space Telescope images. 
Furthermore, \citet{farina17} have found a small and faint \Lya halo with $9$ pkpc extent around the QSO J0305$-$3150 at $z=6.61$ even using a high sensitivity of Multi-Unit Spectroscopic Explorer (MUSE: \citealp{bacon15}).
These studies imply a possible absence of large \Lya halos at $z>6$, which may be attributed to the evolution of the CGM and/or some physical properties of QSOs.

In this paper, we first present new narrow-band observations of the \Lya halo around the QSO \ourq\ at 
{$z=6.4164$ (by [\ion{C}{ii}] $\lambda_\text{rest}=157.741$ $\mu$m: \citealp{willo17}). Previous studies based on spectroscopic data have confirmed the presence of an extended Ly$\alpha$ halo with $>15$ pkpc wide \citep{goto09, goto12, willo11}, but the reported extent and total \Lya luminosity are probably underestimated due to flux loss in spectroscopic data.
In this study, we examine the halo using deep imaging data taken with the custom narrow-band filter $NB906$ on Suprime-Cam which enables us for the first time to map out the whole extension of \ourq\  \Lya halo, and to obtain more accurate measurements on its size and luminosity. 
We then combine our results with all QSOs' \Lya halo data available in the literature ('compilation sample'), to examine halo properties and their correlations with properties of hosting QSOs over $z\sim3-6$. 
Finally, we discuss the evolution of \Lya halos as a function of QSO age at $z>6$, and as a function of redshift over $2<z<6$.
Because we mainly focus on the properties and evolution of \Lya halos, we do not include QSOs whose halo is undetected in our compilation sample except for $z>6$ objects.

The structure of this paper is as follows.
We show Suprime-Cam data and its analysis in Section 2, and results on \ourq\  in Section 3. 
Results from the compilation sample are shown in Section 4. 
We discuss the correlations between \Lya halos' scales (luminosity and size) and physical properties of hosting QSOs, and the evolution with QSO age derived from the correlations seen in the $z>6$ sample in Section 5.1. We also present discussion on the redshift evolution of \Lya halos and the CGMs from the compilation sample in Section 5.2. 
Finally, a summary is given in Section 6. 
Throughout this paper, we use AB magnitudes and adopt a cosmology parameter set of ($\Omega_m$, $\Omega_\lambda$, $H_0$) = ($0.3$, $0.7$, $70$ km s$^{-1}$ Mpc$^{-1}$). In this cosmology, $1$ arcsec corresponds to a transverse size of $5.5$ pkpc at $z=6.4$.

\section{Data and Analysis}
\label{data}

We observed a $0.219$ deg$^2$ field centered on \ourq\  with Suprime-Cam \citep{miya02} on the Subaru Telescope through the $z_r$ and $NB906$ filters \citep{goto09, utsu10,goto17}.
Because details of our observations and data reduction have already been reported \citep{goto09, utsu10,goto17}, we give a brief summary here.

The $z_r$ band is a special broad-band filter constructed by \citet{shimasaku05} with $\lambda_\text{c}= 9853$ \AA\ and FWHM $= 568$ \AA. 
The target field was observed with this filter during an engineering run in 2008 August and UH time in 2009 June, together with the $i'$ and $z'$ bands. The exposure time was $12,532$ sec.
As detailed in \citet{goto09}, the final combined $z_r$ image has a PSF FWHM of $0\farcs62$ and a $3\sigma$ limiting magnitude of $25.46$ mag in a $1\farcs2$ aperture (see also \citealp{goto17}).

The $NB906$ is a custom narrow-band filter developed to investigate the extended \Lya emission of \ourq\  and \Lya emitters around it \citep{goto17}. Its central wavelength and FWHM are $9050$ {\AA} and $158$ {\AA}, respectively.
\citet{willo11} and \citet{goto12} have shown from their spectroscopic observational data that the \ourq's extended \Lya emission in the CGM spans over $9000 < \lambda_\text{obs} < 9100$ {\AA}.
Our custom narrow-band filter fully covers this wavelength range.
We should also mention the fact that different lines can give different systemic redshifts. Recently, a relatively large $\sim1000$ km s$^{-1}$ systemic-redshift offset between [\ion{C}{ii}] and \ion{Mg}{ii} lines has been reported for some high-$z$ QSOs (e.g., \citealp{venemans16}); 
we cannot know which is correct.
For such QSOs, the dedicated narrow-band filter may fail to catch the \Lya emission if it is designed based on the wrong line.
However, since the offset found for \ourq\ is negligible, 
[\ion{C}{ii}] ($z_\text{sys}^{[\ion{C}{ii}]} = 6.4164\pm0.0008$: \citealp{willo17}) and 
\ion{Mg}{ii} $\lambda_\text{rest}=2799.49$ {\AA} ($z_\text{sys}^{\ion{Mg}{ii}} = 6.417\pm0.002$: \citealp{willo10}),
the \Lya emission from \ourq\ is expected to be well within the FWHM of our custom narrow-band filter.

$NB906$ observations were performed in 2015 September with an exposure time of $23,044$ sec.
The images were reduced in the same manner as in \citet{utsu10}. 
After subtracting a bias level estimated by an overscan region and flat fielding, we mean-combined all the images into a single mosaic image. Photometric calibration was performed by comparing stars in a reference field $2$ deg. away to the north. A photometric transformation equation was determined by convolving \citet{gunn83}'s stellar SEDs with the response curves that include both optical and atmospheric transmissions.
The combined image has a PSF FWHM of $0\farcs64$ and a $5\sigma$ limiting magnitude of $25.73$ mag.

We produce a Ly$\alpha$ image in the following two steps.
First, we smooth the $z_r$ and $NB906$ images with Gaussian kernels to match their PSF sizes to the largest among the raw images before stacking, $0\farcs7$ FWHM. 
Second, we subtract the smoothed $z_r$ image from the smoothed $NB906$ image by using the following relation given in the literature (e.g., \citealp{yang09,arrigoni16}):
\begin{equation}
    F_{Ly\alpha} = F_\text{NB} - a \frac{F_{zr}}{\Delta \lambda_{zr}} \Delta \lambda_\text{NB},
\end{equation}
where 
$F_{zr}$ and $F_\text{NB}$ are the fluxes in the $z_r$ and $NB906$ bands, and $F_{Ly\alpha}$ the flux of the \Lya line; 
$\Delta \lambda_{zr}$ and $\Delta \lambda_\text{NB}$ are the FWHMs of the $z_r$ and $NB906$ bands;
and $a$ is a parameter evaluated from the spectral index $\alpha$.
Because we adopt $\alpha_\nu = -0.5$\footnote{Here we define $f_\nu \propto \nu^{\alpha_\nu}$} \citep{willo07}, $a$ is estimated as 0.96. 
The SB limit of the Ly$\alpha$ image thus obtained in a $1\arcsec$ aperture is $6.7\times10^{-18}$ erg s$^{-1}$ cm$^{-2}$ arcsec$^{-2}$ at the $3\sigma$ level.
We also stack $100$ bright stars in our narrow-band image to produce a PSF image.

%
%
\subsection{The subtraction of the PSF component}
To examine whether \ourq\ has an extended \Lya halo, we remove the contribution from the PSF (i.e., point-source) component of the central QSO. 
We conduct the PSF subtraction by the following three approaches to find the best way to estimate the halo flux and size.

\begin{itemize}
	\item[1)] One is to use the PSF image obtained from $100$ stars in the $NB906$ image.
	First, we scale the PSF image to match the total flux within a $1\arcsec$ radius to that of the \Lya image of \ourq. Then, we subtract the scaled PSF image from the \ourq\  image, and obtain a residual image ($residual\ image\ 1$). We confirm that \Lya emission extends over $\sim2\arcsec$ in radius with $>3\sigma$ SB levels. 
	However, there is a hole at the position of \ourq\  which is probably caused by over-subtraction of the PSF component of the central QSO.
	\item[2)] The second approach is to use a 2-D Gaussian model in order to minimize the effect of over-subtraction. 
    First, we produce a 2-D Gaussian model image by fitting the PSF image generated from the $100$ stars. 
    Then, we scale this Gaussian model image and subtract it from the \ourq\  image in the same manner as in approach 1). The residual image thus obtained is referred to as $residual\ image\ 2$. The presence of extended \Lya emission over $\sim1\arcsec$ is confirmed also in this image. While a hole is still seen, its size is much smaller than that in $residual\ image\ 1$, suggesting that the over-subtracted flux has been largely recovered.
	\item[3)] The third approach is to use a 2-D Moffat model image.
	The SB profile and hence the \Lya luminosity of the halo evaluated from $residual\ image\ 2$ may slightly overestimate the true size and luminosity, because the real PSF profile is known to extend more than a Gaussian profile at large radii albeit with very low amplitudes (e.g., \citealp{king71,racin96,bers07}). Thus we perform a PSF subtraction with a model Moffat image \citep{moffat69} which can describe well the extended profile of the PSF. 
	The methodology is the same as approach 2). We first make a 2-D Moffat image by fitting the PSF image obtained from the $100$ stars.
	Then we scale the Moffat model image, and subtract it from the \ourq\ image, which is referred to as $residual\ image\ 3$. 
\end{itemize}
We show the results from the above three approaches in Figure \ref{fig:radip} (a).
At $1\arcsec<r<2\arcsec$, the shape and brightness of the SB profiles obtained from the three approaches show a large difference. However, they become consistent within the error bars at $2\arcsec<r<4\arcsec$.
Because both PSF models (Gaussian and Moffat) reproduce the observed PSF similarly well, we adopt the mean of $residual\ image\ 2$ and $residual\ image\ 3$ 
(produced in approaches 2) and 3))
as better representing the true \Lya emission of the halo, and use this image and SB profile in the following analysis.
We show the PSF-subtracted \Lya image of \ourq\ in Figure \ref{fig:img}, and the \Lya SB profile of \ourq\ and the PSF profile in Figure \ref{fig:radip} (a).
Figure \ref{fig:radip} (b) and (c) also compare the tangentially averaged SB profile of \ourq's \Lya halo with those of low-$z$ QSOs and a $z\sim 6$ QSO obtained by previous studies \citep{borisova16b,farina17,arrigoni18b}.
Note that the SB profile of \ourq\ is calculated directly from the data.

\begin{figure}
	\includegraphics[width=\linewidth]{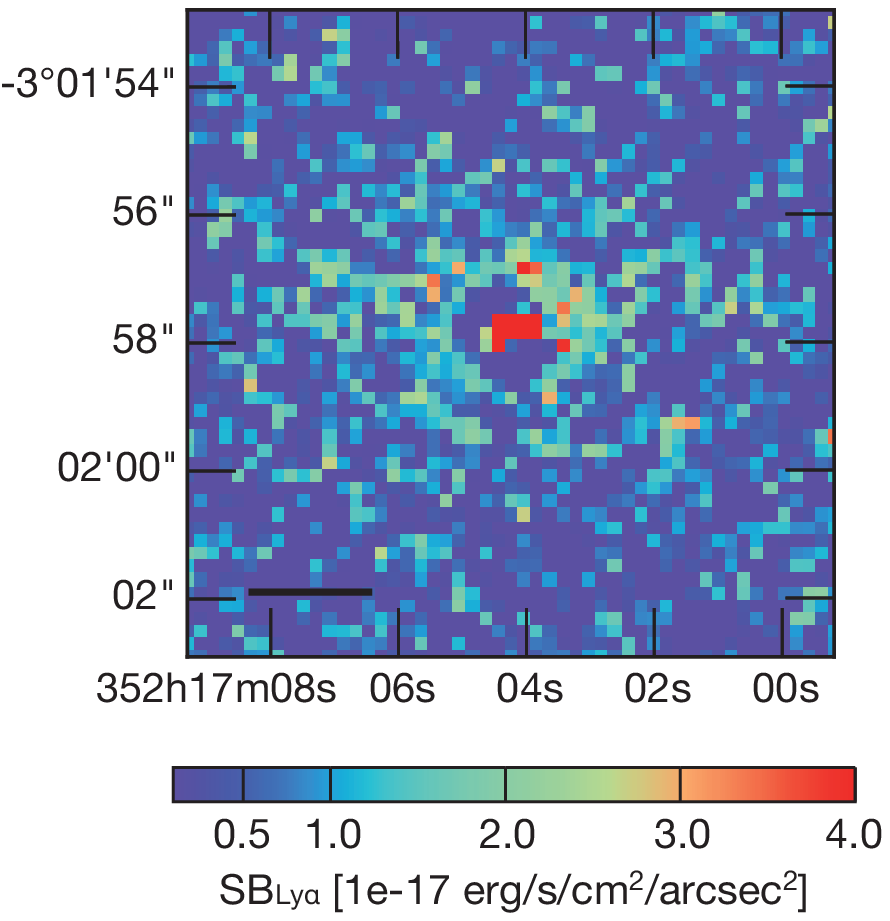} 
	\caption{
	The \Lya image of \ourq\ is presented. 
	}
	\label{fig:img}
\end{figure}

\begin{figure*}
	\begin{center}
	\includegraphics[width=\linewidth]{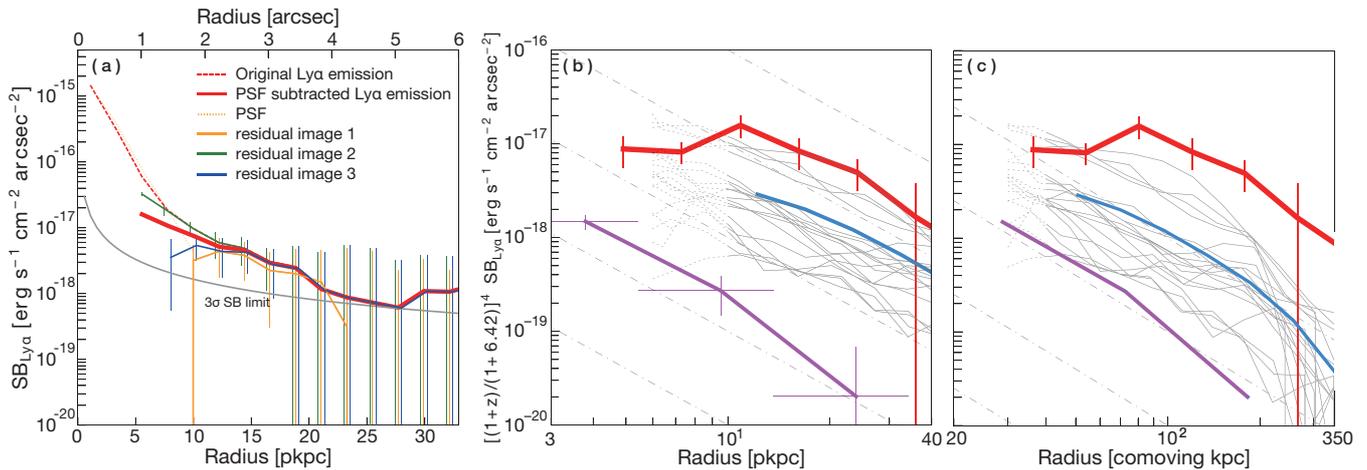} 
	\caption{
	(a) Tangentially averaged radial SB profile of the \Lya emission of \ourq\  
	together with $residual\ images$, measured in $0\farcs4$ widths. Red dashed and solid lines represent, respectively, the \Lya emission before subtraction of the QSO component and the finally adopted result in this study (i.e., mean of $residual\ images\ 2$ and $3$). 
	Yellow, green and blue solid lines indicate $residual\ images\ 1$, $2$, and $3$, respectively. 
	The SB profile of the PSF obtained from $NB$ data is also plotted with a yellow dotted line.
	(b) PSF-subtracted, redshift-dimming corrected SB profile of \ourq\  (red) together with those of radio-quiet QSOs at $z\sim3$ (\citealp{borisova16b}; black),	the mean of the QSO MUSEUM (\citealp{arrigoni18b}; blue) and J0305-3150 at $z=6.61$ (\citealp{farina17}; purple).
	The SB of \ourq\  is measured up to 
	$r=6\farcs5$
	($36$ pkpc) in annuli of $\Delta \log r=0.17$ pkpc corresponding to $0\farcs2-2\farcs2$. Since \citet{borisova16b} have regarded only \Lya emission outside of $r=10$ pkpc as the halo component, we plot their SB profiles at $r<10$ pkpc with dashed lines. The five dash-dotted lines, plotted for reference, are SB$(r) \propto r^{-2}$ power laws with different amplitudes of $10^{-14}$, $10^{-15}$, $10^{-16}$, $10^{-17}$, and $10^{-18}$ erg s$^{-1}$ cm$^{-2}$ arcsec$^{-2}$ from top to bottom. 
	(c) Same as panel (b) but plotted in comoving scale. (A color figure is available in the online journal.)}
	\label{fig:radip}
	\end{center}
\end{figure*}

\section{Results from Suprime-Cam data}
\label{sec:results}

\subsection{The luminosity and spatial extent of \ourq's  \Lya halo}
\label{sec:result_2329LAH}

We find that the \Lya emission of \ourq\  is more extended than the PSF (Figure \ref{fig:radip} (a)) as well as than the continuum emission (Figure \ref{fig:img}), thus confirming the presence of a \Lya halo. 
We use the SB profile (see also Figure \ref{fig:radip}) to measure the spatial extent of the \Lya halo, i.e., the largest diameter where the SB is above the SB limit, to be $6\farcs8$, or $37$ pkpc. The total \Lya luminosity within an annulus of radii from $1\arcsec$ to 
$3\farcs4$
is $9.8 \pm 2.3 \times10^{43}$ erg s$^{-1}$. The \Lya luminosity of the QSO is $6.6 \times 10^{44}$ erg s$^{-1}$.

The \Lya halo of this QSO has also been investigated in previous studies. 
\citet{goto09} have indicated a possible presence of $4\arcsec$-wide \Lya halo. Spectroscopic follow-up observations by the same group have obtained the total \Lya luminosities of the halo and the hosting QSO to be
$1.7\times10^{43}$ erg s$^{-1}$ and $6.2\times10^{44}$ erg s$^{-1}$, respectively \citep{goto12}. 
On the basis of spectroscopic data, \citet{willo11} have also shown the presence of a \Lya halo over $15$ pkpc ($\sim 2\farcs7$) with a \Lya luminosity of $8\times10^{43}$ erg s$^{-1}$.
The \Lya halo found in this study is a factor of $>1.7$ more extended, and a factor of $>1.9$ more luminous than those obtained by the previous studies. These discrepancies likely result from differences in the methods 
to measure halo luminosities.
The spectroscopic data could also underestimate the total flux of the halo as well as its extent because of finite slit widths and/or fixed slit directions \citep{willo11,goto12}.

\subsection{Comparison of the halo SB profile between \ourq\  and other QSOs}
\label{sec:result_2329SBp}

We compare the SB profile of \ourq's halo with those of other QSOs' in Figure \ref{fig:radip} (b) and (c). Black lines represent bright radio-quiet QSOs at $z\sim3$ \citep{borisova16b} and a blue line is the QSO J0305$-$3150 at $z=6.61$ \citep{farina17}; the effect of SB dimming has been corrected for all objects. 
The profiles of \citet{borisova16b} QSOs within $r=10$ pkpc are plotted as dotted lines.

A comparison with J0305$-$3150 shows that \ourq's halo has an about $10$ times brighter SB profile than J0305$-$3150's, albeit with similar profile slopes. As a result, J0305$-$3150's halo has a factor $50$ fainter total \Lya luminosity and a factor of $4.2$ smaller size than \ourq's, despite a deeper SB limit of $1.1\times10^{-18}$ erg s$^{-1}$ cm$^{-2}$ arcsec$^{-2}$ at $3\sigma$ level \citep{farina17}.
The SB profile of \ourq's halo lies at the bright end of the distribution of $z\sim3$ halos.
We estimate the power-law slope of \ourq's SB profile over $10$ pkpc to $24$ pkpc by fitting the formula SB$=C_p r^\alpha$, where $C_p$ is a normalization parameter and $\alpha$ the slope of the power law. We obtain (log$_{10}$($C_p$), $\alpha$) = ($-15.08\pm0.47$, $-1.82\pm0.35$) in physical scales and ($-13.49\pm0.78$, $-1.82\pm0.35$) in comoving scales. 
From a survey of $z=3$ \Lya halos by the MUSE, \citet{borisova16b} and \citet{arrigoni18b} have reported the power-law slope of the mean SB profile as $\alpha=-1.8$ and $-1.96$, respectively. 
Although the fitting range of \ourq\ is narrower than those of the $z=3$ \Lya halos, the power-law slope of \ourq\ is consistent with their measurements.
These comparisons suggest that \Lya halos of QSOs can be modeled by profiles with a common shape irrespective of redshift.

We also compare those SB profiles in comoving units by scaling their radii with ($1+z$) in Figure \ref{fig:radip} (c). 
First, we find that the difference between J0305$-$3150 and QSOs from \citet{borisova16b} seen in Figure \ref{fig:radip} (b) is clearly reduced. The SB profile of J0305$-$3150 comes into overlap with the faint end of \citet{borisova16b}'s sample, although the SB profile of \ourq\  is shifted slightly above the bright end.
A similarity of SB profiles in comoving units has also been demonstrated by \citet{ginolfi18}. They have found that the sizes of \Lya halos at $z\sim5$ become comparable with those at $z\sim3$ when they have accounted for the cosmological growth of dark matter halos by scaling by ($1+z$) \citep{barkana01}. 
Detailed discussion will be presented in Section \ref{sec:M_DM}.

\begin{table}
	\caption{The parameters and the number of objects plotted in individual figures.}
	\label{tab:sampleinfo}
	\begin{tabular}{lccccccc}
 	\hline
 		Figure number		& $N_\text{sample}$ & Spearman's $\rho$	\\
	\hline \hline
		Figure \ref{fig:QSO_z} $L_{Ly\alpha}$ vs redshift 
							& 135	& --	\\
	\hline
		Figure \ref{fig:QSO_z} $L_\text{Bol}$ vs redshit
							& 78	& --	\\
	\hline
		Figure \ref{fig:QSO_z} $L_{Ly\alpha}$/$L_\text{Bol}$ vs redshift
							& 71	& --	\\
	\hline
		Figure \ref{fig:QSO_z} $d$ vs redshift cyan points
							& 109 	& --	\\
							& (144)	& --	\\
	\hline
		Figure \ref{fig:QSO_host} $L_{Ly\alpha}$ vs $d$ painted points
							& 106	& $0.75\pm0.05$	 \\
							& (133)	& ($0.57\pm0.07$)	\\
	\hline
		Figure \ref{fig:QSO_host} $M_\text{BH}$ vs $d$ painted points
							& 59	& $-0.28\pm0.13$	\\
							& (72)	& ($-0.16\pm0.13$)	\\
	\hline
		Figure \ref{fig:QSO_host} $M_\text{BH}$ vs $L_{Ly\alpha}$/$L_\text{Bol}$ 
							& 65	& $-0.44\pm0.11$	\\
	\hline
		Figure \ref{fig:QSO_dust} $M_\text{BH}$ vs IRX
							& 9	& $0.12\pm0.43$	\\
	\hline
		Figure \ref{fig:QSO_dust} $d$ vs IRX
							& 5	& $-0.50\pm0.52$	\\
							& (7) & ($-0.04\pm0.52$)	\\
	\hline
		Figure \ref{fig:QSO_dust} $L_{Ly\alpha}$/$L_\text{Bol}$ vs IRX
							& 8	& $-0.54\pm0.43$	\\
	\hline
		Figure \ref{fig:SB10} SB$_{Ly\alpha}^{r=10}$ vs $z$
							& 124	& --	\\  
	\hline
 	\end{tabular}
	\begin{tablenotes}
		\small
		\item The number in parentheses in $N_\text{sample}$ column is the number of objects with $d_\text{org}$ measurements. 
		The standard deviation of Spearman's $\rho$ derived from $1000$ times resampling is regarded as the error in the $\rho$ estimate. 
	\end{tablenotes}
\end{table}
%

\begin{figure*}
	\begin{center}
	\includegraphics[width=\linewidth]{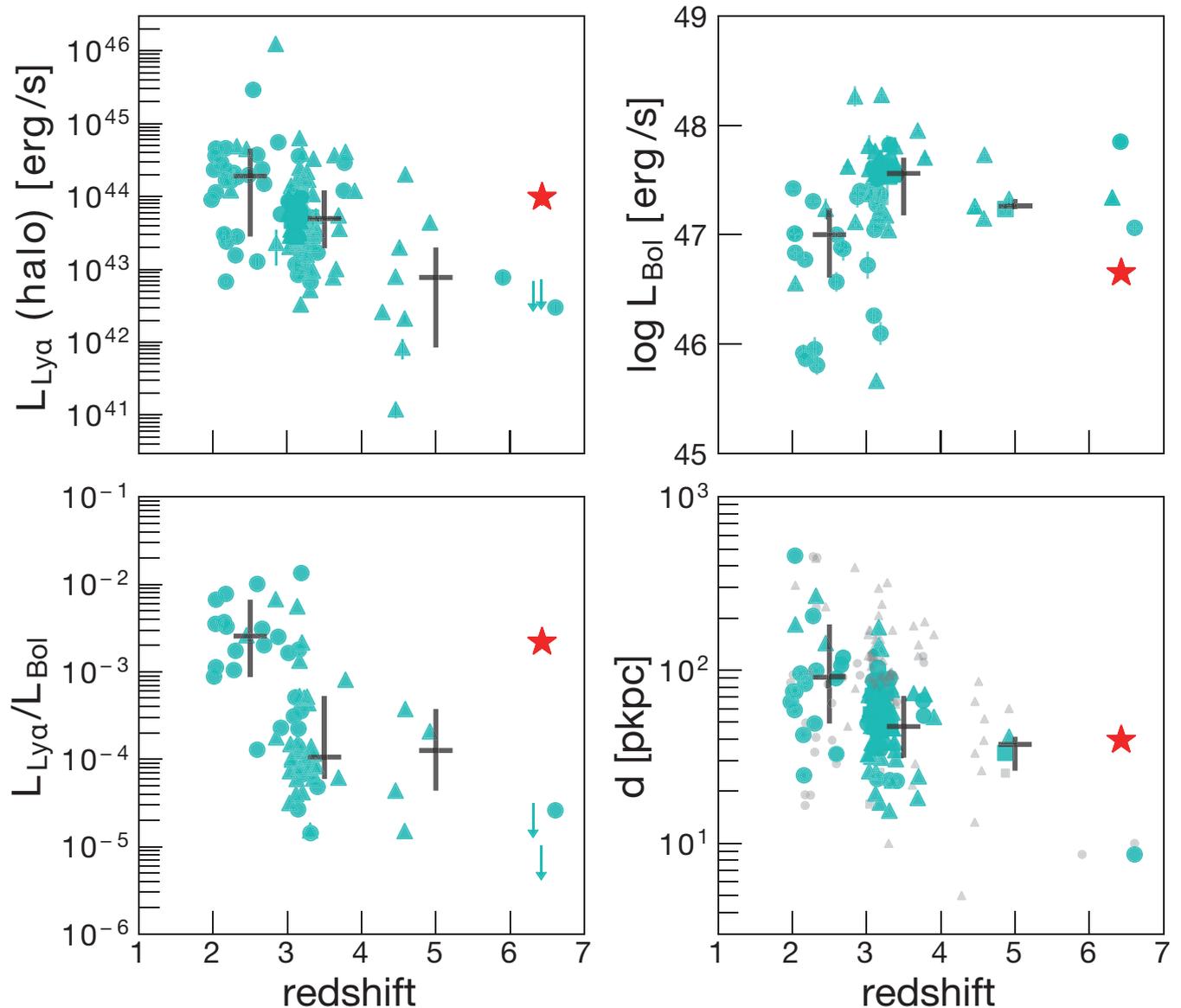} 
	\caption{\Lya luminosity, $L_\text{Bol}$, $L_{Ly\alpha}$/$L_\text{Bol}$, and $d$, as a function of redshift.  
	\ourq\  is plotted in red, while the other QSOs in cyan. Circle and triangle symbols represent radio-loud and radio-quiet QSOs, respectively; those without type information are shown by squares. Median values at $z=2$, $3$, and $4-5$ bins are overlaid as black crosses.  
	In the top left and bottom left panels, upper limits of the \Lya luminosity for two undetected $z>6$ \Lya halos are also plotted \citep{deca12}.
	In the bottom right panel, originally estimated sizes in the literature ($d_\text{org}$) are also plotted with grey symbols.
	(A color figure is available in the online journal.)
	}
	\label{fig:QSO_z}
	\end{center}
\end{figure*}

\begin{figure*}
	\begin{center}
	\includegraphics[width=\linewidth]{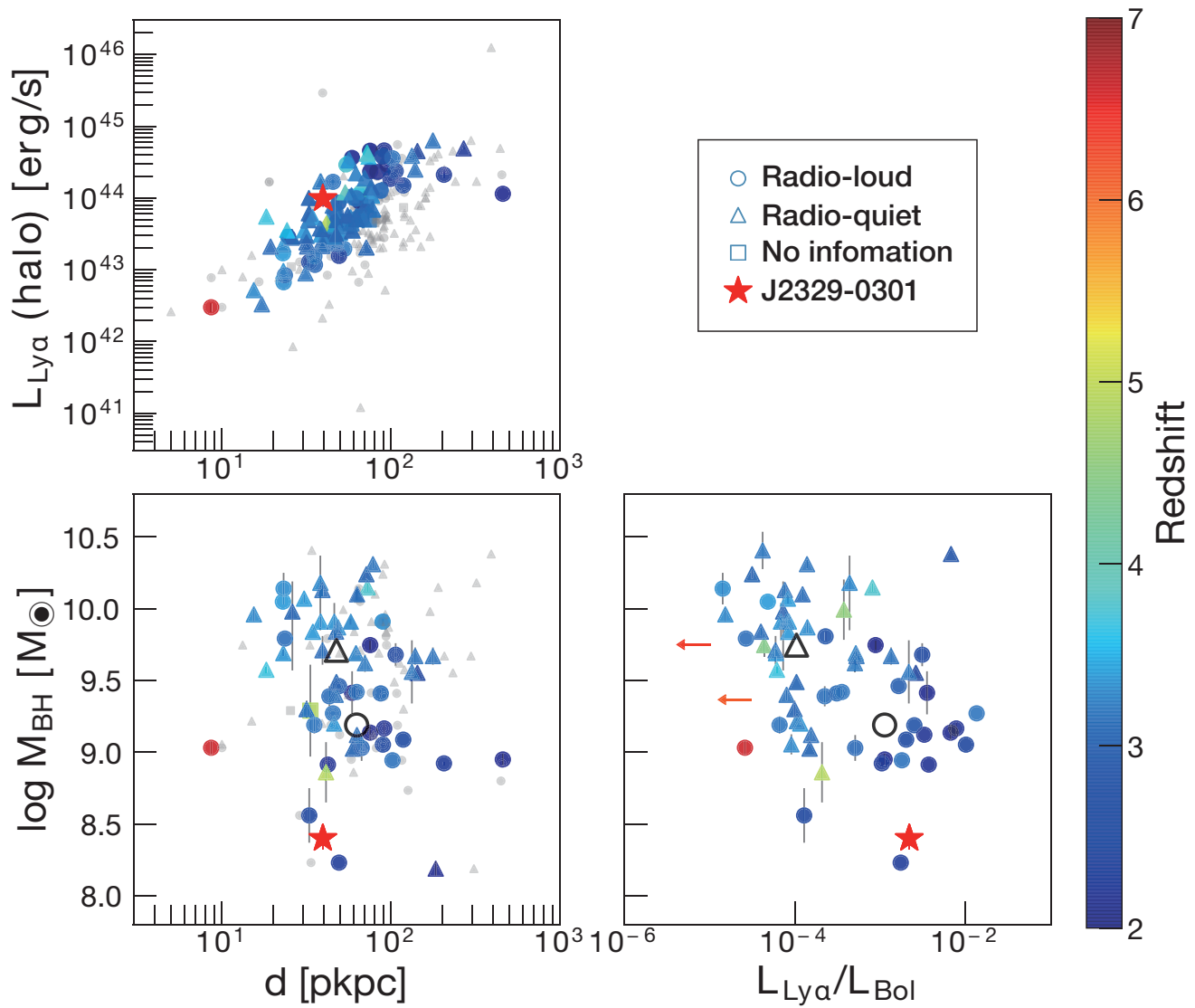} 
	\caption{Correlations of $d$ and $L_{Ly\alpha}$/$L_\text{Bol}$ with the $L_{Ly\alpha}$ and $M_\text{BH}$ of hosting QSOs. \ourq\  is shown as a star. Circle and triangle symbols represent radio-loud and radio-quiet QSOs, respectively; those without information are shown by squares. All symbols except grey ones are color-coded according to redshift. The $L_{Ly\alpha}$/$L_\text{Bol}$ of J1030$+$0524 and J1148$+$5251 from \citet{deca12} is an upper limit. 
	(Left) $L_{Ly\alpha}$ and $M_\text{BH}$ as a function of $d$. Original $d$ estimates taken from the literature are also plotted with grey symbols. 
	(Right) $M_\text{BH}$ plotted against $L_{Ly\alpha}$/$L_\text{Bol}$. 
	(Bottom) Median values of the radio-loud and quiet samples are also displayed by a void circle and a void triangle.
	(A color figure is available in the online journal.)
	}
	\label{fig:QSO_host}
	\end{center}
\end{figure*}

\begin{figure*}
	\begin{center}
	\includegraphics[width=\linewidth]{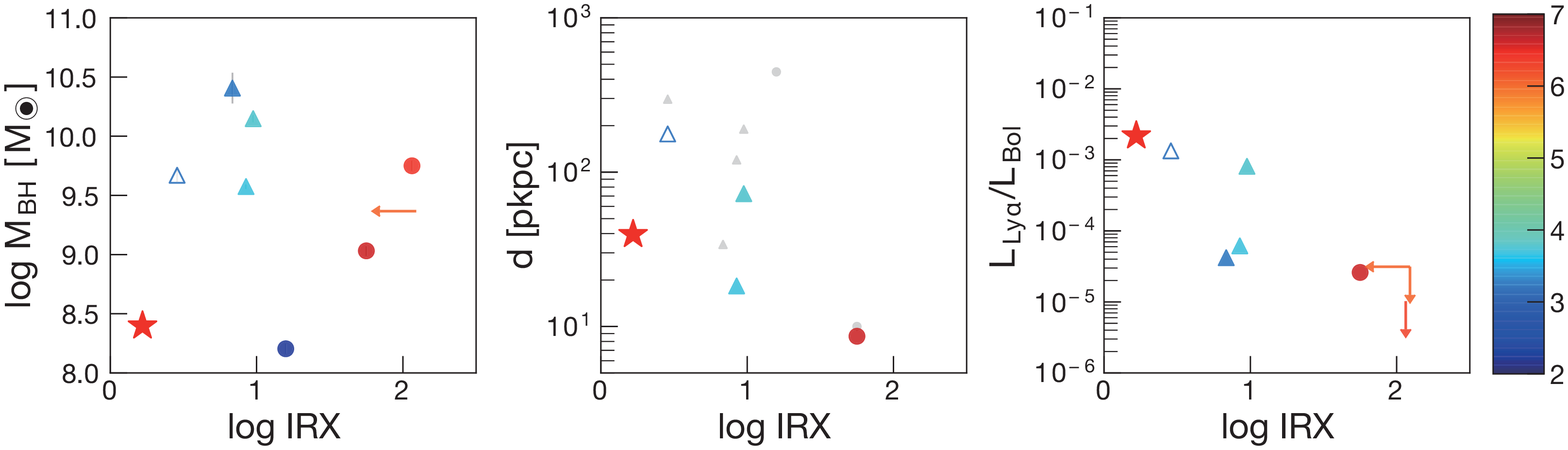} 
	\caption{
	Correlations of IRX with $M_\text{BH}$, $d$ and $L_{Ly\alpha}$/$L_\text{Bol}$. 
	The meanings of symbols and colors are the same as in Figure \ref{fig:QSO_host}.
	The IRX value of a QSO shown by a white triangle is derived using a FIR luminosity instead of a total IR luminosity.
	(A color figure is available in the online journal.)
	}
	\label{fig:QSO_dust}
	\end{center}
\end{figure*}


\section{Results from the compilation sample: 
dependence of \Lya halo properties on redshift and QSO properties
}
\label{sec:result_comp}

We plot
in Figures \ref{fig:QSO_z}, \ref{fig:QSO_host}, and \ref{fig:QSO_dust} three parameters characterizing \Lya halos (spatial extent $d$ [pkpc], total \Lya luminosity $L_{Ly\alpha}$ [erg s$^{-1}$], and the ratio of the halo \Lya luminosity to the hosting QSO's bolometric luminosity $L_{Ly\alpha}$/$L_\text{Bol}$) against
[1] redshift,
[2] two parameters associated with the supermassive black hole of the hosting QSO (black hole mass $M_\text{BH}$ [M$_\odot$] and the Eddington ratio) and $L_\text{Bol}$, and
[3] the dustiness of the hosting QSO (the UV-optical spectral index $\alpha_\nu$, and the IR-to-UV luminosity ratio IRX $=L_\text{IR}$/$L_\text{UV}$).
We have collected all available data from the literature; see footnote.\footnote{
Data of QSOs \Lya halos are from
\citet{stei91,heckman91a,heckman91b,bermer92,roett97,vanO97,leh98,leh99,berg99,fyn00,bunker03,weid04,chris06,francis06,cour08,barrio08,smith09,yang09,matsu11,north12,deca12,hump13,rauch13,cantalupo14,roche14,martin14,husband15,henna15,fumag16,borisova16b,fath16,north17,bayiss17,kikuta17,farina17,cai17,Cai_2018,ginolfi18,arrigoni18b,arrigoni18}.
Other data of $M_\text{BH}$, Eddington ratio, $\alpha$, UV luminosity and FIR or total IR luminosity are from
\citet{hughes97,penter03,vest03,vest09,willo07,shen11,weed12,cal14,leips14,paris14,paris17,derosa14,ma15,tsai15,bana16,murphy16,venemans16,koz17,mazzu17}.
We correct both $L_{Ly\alpha}$ and $L_\text{bol}$ based on the adopted cosmology in this study.
}
Below we explain each parameter in detail.
\begin{itemize}
	\item {\bf Spatial extent $d$, $d_\text{org}$:} 
		We virtually put all objects at $z=6.61$ and measure their extents or diameters, at a common SB level of $1\times10^{-18}$ erg s$^{-1}$ cm$^{-2}$ arcsec$^{-2}$ at $z=6.61$.\footnote{
It means an SB limit of [($1+6.61$)$^4$/($1+z_{sys}$)$^4$]$\times1\times10^{-18}$ erg s$^{-1}$ cm$^{-2}$ arcsec$^{-2}$.
}
		For reference, we also plot original extents taken from the literature with grey-colored symbols in the same panels. In the following sections, $d$ denotes an anew measured extent and $d_\text{org}$ an original one.
	\item {\bf Total \Lya luminosity $L_{Ly\alpha}$:} 
		While an ideal definition of the total luminosity would be the one enclosed within the extent $d$, such measurements are available only for a small  fraction of the sample. Thus, we use \Lya luminosities presented in the literature. For the two $z>6$ QSOs whose halo is undetected (J1030$+$0524 and J1148$+$5251, \citealp{deca12}), we show upper limits derived by assuming a $10$ pkpc extent and an average $5\sigma$ SB value of $1\times10^{-17}$ erg s$^{-1}$ cm$^{-2}$ arcsec$^{-2}$, following the procedure by \citet{farina17}.
	\item {\bf Ratio of the total \Lya luminosity to the hosting QSO' bolometric luminosity, $L_{Ly\alpha}$/$L_\text{Bol}$:} 
		We define this parameter to characterize the relative prominence of a halo.
	\item {\bf $M_\text{BH}$ and $L_\text{Bol}$:}
		Black hole masses $M_\text{BH}$, bolometric luminosities $L_\text{Bol}$, and Eddington ratios are taken from the literature if available;  
		$M_\text{BH}$ have been estimated using black hole mass scaling relations with \ion{Mg}{ii}, \ion{C}{iv} and \ion{H}{$\beta$} luminosities; 
		$L_\text{Bol}$ have been determined from spectral energy distributions (SEDs) from X-ray to radio wavelengths, or from monochromatic luminosities at $5100$, $3000$, or $1350$ {\AA} with bolometric corrections \citep{richard06,shen08}.
	\item {\bf Spectral index $\alpha_\nu$:} 
		Taken from the literature if available. They have been derived by fitting a single power law of $f_\nu \propto 
		\nu^{\alpha_\nu}$ 
		over $1000-3000$ or $1000-10000$ {\AA} in the rest-frame \citep{penter03,willo07,leips14,paris14,paris17,mazzu17}. 
	\item {\bf IR to UV luminosity ratio, IRX:} \
		This ratio, called the IR excess, is an indicator of the dustiness of a galaxy. We estimate IRX values with UV luminosities derived from absolute magnitudes at $1450$ or $1550$ {\AA} and total IR luminosities given in the literature. If a total IR luminosity is unavailable, a total far-IR luminosity is used instead. Due to few radio studies in our compilation sample, the number of objects with IRX estimates is limited.
\end{itemize}

Unfortunately, not all objects have a full set of measurements except for $d_\text{org}$ and redshift. For this reason, different subsamples are used in different plots, as summarized in Table \ref{tab:sampleinfo}. 
Also shown in this table is Spearman's rank correlation coefficient $\rho$ for each plot in Figures \ref{fig:QSO_host} and \ref{fig:QSO_dust}.
The standard deviation of $\rho$ obtained from $1000$ times data resampling is considered as its error.

\subsection{Redshift dependence of halo properties}
\label{comp_z}

Figure \ref{fig:QSO_z} shows the $L_{Ly\alpha}$, $L_\text{Bol}$, $L_{Ly\alpha}$/$L_\text{Bol}$ and $d$ of \Lya halos as a function of redshift. The median values at $z=2$, $3$, and $4-5$ are also plotted with an error bar for reference.
We find a gradual decline in $L_{Ly\alpha}$, $L_{Ly\alpha}$/$L_\text{Bol}$ and $d$ with redshift, while $L_\text{Bol}$ does not show such a decline.
In addition, the decrease in the spatial extent is notable, which is found in both $d$ and  $d_\text{org}$. 
Interestingly, \ourq's halo has an spatial extent as small as average $z\sim2-3$ halos despite having relatively high $L_\text{Bol}$ and $L_{Ly\alpha}$/$L_\text{Bol}$ values.

The observed declines of $L_{Ly\alpha}$, $L_{Ly\alpha}$/$L_\text{Bol}$ and $d$ with redshift may be partly artificial.
Different kinds of observations (imaging or spectroscopy), different SB limits, and different target selections among the studies may induce artificial trends.
In fact, some previous studies have found similar declines of $d_\text{org}$ and/or $L_{Ly\alpha}$ with redshift \citep{farina17,ginolfi18}, whereas others have not \citep{fath16}. 
Nevertheless, Figure \ref{fig:QSO_z} clearly demonstrates a systematic trend that bright ($>$ several $\times10^{44}$ erg s$^{-1}$) and extended ($>100$ pkpc) \Lya halos are absent at $z>4$. 
Considering the facts that the average $L_\text{Bol}$ is unchanged or even increases with redshift, and that the bolometric luminosity is proportional to the ionizing luminosity in the first order (e.g., \citealp{lu99}), bright and extended \Lya halos such as found in $z=2-3$ (e.g., \citealp{cantalupo14, martin14, henna15, borisova16b,cai17,arrigoni18}) should be detected even at $z>4$ if there is no evolution. 
This striking result indicates some redshift evolution of \Lya halos. 
We will discuss this matter in Section \ref{sec:CGM_QSO}.

\subsection{Correlations with properties of hosting QSOs}
\label{comp_prop}

We find no significant correlation between halo properties and QSOs' $L_{Ly\alpha}$, $L_{Ly\alpha}$/$L_\text{Bol}$, $M_\text{BH}$ and Eddington ratio except in $d$ vs $L_{Ly\alpha}$.
Thus, we only display $d$ vs $L_{Ly\alpha}$ and two plots showing very weak correlations in Figure \ref{fig:QSO_host}.

A positive and strong correlation is seen between $d$ and $L_{Ly\alpha}$ (top left plot of Figure \ref{fig:QSO_host}), with a Spearman's coefficient of 
$\rho=0.75\pm0.05$ ($0.57\pm0.07$)
if $d_\text{org}$ is used instead.\footnote{
We regard the standard deviation of Spearman's rank correlation $\rho$ obtained from $1000$ times resampling of data points as $1\sigma$ error.
}
In fact, such a positive correlation is naturally expected from the finding in Section \ref{sec:result_2329SBp} that \Lya halos have similar SB profiles. We derive the best-fit linear regressions in the log-log plot of 
$\log L_{Ly\alpha}$ [erg s$^{-1}$] $= (41.53\pm0.11) + (1.32\pm0.06) \times \log d$ [pkpc] and 
$\log L_{Ly\alpha} = (43.79\pm0.21) + (0.25\pm0.12) \times \log d_\text{org}$. 
A similar correlation has also been reported in the literature \citep{fath16,north17}. 
\citet{fath16} have found the best-fit power law of the form $d_\text{org} \propto L_{Ly\alpha}^{b}$ with $b=0.5$ for \citet{chris06}'s sample and $b=0.7$ for their own sample, corresponding to a slope of $2$ and $1.42$ in the $L_{Ly\alpha}$ vs $\log d_\text{org}$ plot. 

No apparent correlation is found between $d$ and $M_\text{BH}$ with Spearman's $\rho$ of $\rho=-0.28\pm0.13$ ($\rho=-0.16\pm0.13$ for $d_\text{org}$), while a moderate correlation is seen between $L_{Ly\alpha}$/$L_\text{Bol}$ and $M_\text{BH}$ with $\rho=0.44\pm0.11$ (bottom two panels in Figure \ref{fig:QSO_host}). 
If limited to the $z>6$ QSOs colored in red, we find a possible negative trend in both panels, despite a very small sample size, that QSOs with a larger $M_\text{BH}$ tend to possess a larger and brighter \Lya halo.
If larger $M_\text{BH}$ means a longer elapsed time, the weak decrease in $L_{Ly\alpha}$/$L_\text{Bol}$ with $M_\text{BH}$ found among the $z>6$ QSOs may indicate that younger QSOs possess more luminous (and possibly more extended; lower left panel of Figure \ref{fig:QSO_host}) \Lya halos.
We discuss this matter in Section \ref{sec:life_QSO}.

The $L_{Ly\alpha}$/$L_\text{Bol}$ vs $M_\text{BH}$ distribution also shows another tendency depending on radio loudness. We display the median values of $L_{Ly\alpha}$/$L_\text{Bol}$ and $M_\text{BH}$ for the radio-loud and quiet samples with void circle and triangle symbols, respectively, in the bottom two panels of Figure \ref{fig:QSO_host}. 
Radio-loud QSOs are found to have more luminous \Lya halos.
This may be due to radio jets associated with radio-loud QSOs, which make it easy for ionizing photons to escape out to the CGM.

\subsection{$M_\text{BH}$, $d$ and $L_{Ly\alpha}$/$L_\text{Bol}$ vs the dustiness of QSOs}
\label{sec:QSO_dustiness}

Finally, we examine the correlations of $M_\text{BH}$, $d$ and $L_{Ly\alpha}$/$L_\text{Bol}$ with $\alpha_\nu$ and IRX.
The spectral index $\alpha_\nu$ is the power-law index of spectra, with larger values meaning bluer spectra.
The IRX is an indicator of dustiness, with higher values meaning higher dust obscuration.

We do not find any significant correlation with $\alpha_\nu$, with small Spearman's $\rho$ of 
$0.32, 0.05$ and $-0.23$ for $M_\text{BH}$, $d$ and $L_{Ly\alpha}$/$L_\text{Bol}$.
On the other hand,
moderate negative correlations are indicated in the plots of $d$ and $L_{Ly\alpha}$/$L_\text{Bol}$ against IRX with
$\rho=(0.12\pm0.43, -0.50\pm0.52, -0.54\pm0.43)$ for ($M_\text{BH}$, $d$, $L_{Ly\alpha}$/$L_\text{Bol}$), although the errors are large due to small sample sizes (Figure \ref{fig:QSO_dust}).
Similarly, the QSOs at $z>6$ colored in red appear to have a trend that those with larger IRX values possess more massive black holes but smaller halo sizes and fainter halo luminosities.
We discuss implications of these trends in Section \ref{sec:life_QSO}.

\begin{table*}
	\caption{Physical parameters of QSOs at $z>6$}
	\begin{threeparttable}[t]
	\label{tab:QSOinfo}
	\begin{tabular}{lrrrr}
 	\hline
 									& \ourq		& J0305$-$3150 	&  J1030$+$0524 & J1148$+$5251 	\\ 
 	\hline
		\Lya halo 					& detection	& detection & non-detection & non-detection	\\
	\hline
		$\log M_\text{BH}$ [M$_\odot$]	& $8.40$	& $8.98-9.08$	& $9.00-9.56$	& $9.75$	\\
		$\log L_\text{Bol}$ [erg s$^{-1}$]	& $46.60$	& $47.06$		& $47.37$		& $47.85$	\\
		Eddington ratio				& 1.3		& $0.68-0.74$	& $0.50$		& $1.01$	\\
		SFR [M$_\odot$ yr$^{-1}$]		& $13$		& $545$			& $< 3165$		& $3801-6000$	\\
		$\log L_\text{IR}$ [L$_\odot$]		& $10.95$	& $12.83$		& $< 13.53$		& $13.74$	\\
		$\log M_\text{dust}$ [M$_\odot$]	& $< 7.00$	& $8.65-9.38$	& $< 8.63$		& $8.85$	\\
 	\hline
	 	References					& 1, 2, 3, 4	& 5, 6			& 4, 6, 7, 8, 9  	& 4, 8, 9, 10	\\
	\hline
 	\end{tabular}
	\begin{tablenotes}
		\small
		\item Note. References are (1) \citealp{willo10}, (2) \citealp{willo17}, (3) \citealp{deca18}, (4) \citealp{cal14}, (5) \citealp{derosa14}, (6) \citealp{venemans16}, (7) \citealp{jiang07}, (8) \citealp{jiang06}, (9) \citealp{lyu16}, and (10) \citealp{leips14}.
	\end{tablenotes}
	\end{threeparttable}
\end{table*}

\section{Discussion}
\label{sec:discussions}

\subsection{Coevolution of QSOs and their \Lya halos}
\label{sec:life_QSO}

\subsubsection{Correlation between \Lya halo properties and the dustiness of QSOs at $z>6$}

Among the four $z>6$ QSOs with \Lya halo data, only \ourq\  has a distinguishing \Lya halo.
In this subsection, we discuss what makes this difference by examining the physical properties of the four $z>6$ QSOs presented in Table \ref{tab:QSOinfo} with help of the results obtained in Section \ref{sec:QSO_dustiness}.

We find from Table \ref{tab:QSOinfo} that \ourq\  has the lowest IR luminosity and the lowest dust mass among the four, while the other three QSOs are luminous in IR or possessing a large amount of dust, implying a close relation between \Lya halos and the amount of dust. 
We also find in Section \ref{sec:QSO_dustiness} that less dusty QSOs with larger $\alpha$ and/or smaller IRX values tend to have more extended and bright \Lya halos.
The relation between dust content and the presence of \Lya halos has also been discussed in the literature. 
Although not for QSOs but LAEs, \citet{hayes13} have found an anti-correlation between the relative extent of the \Lya halo to the star-forming disk and indicators of dust content, indicating that a low dust content is required to produce a extended \Lya halo.
Moreover, \citet{willo11,willo13} have attributed the significant \Lya halo of \ourq\  to the non-detection of dust continuum emission, because the \Lya halo of \ourq\  could be generated by ionizing photons that easily escape to the CGM due to a small dust content. According to their arguments, \ourq\  is in a rare phase that the QSO feedback effectively shuts down the star formation activity of the host galaxy, resulting in a small dust content. 
Our findings from Table \ref{tab:QSOinfo} and Section \ref{sec:QSO_dustiness} support these suggestions, although based on only four objects.
Therefore, the amount of dust probably determines the presence or absence of a \Lya halo.

We also notice that the other parameters of \ourq\  ($M_\text{BH}$, $L_\text{Bol}$ and star-formation rate SFR) are also lower than those of the remaining three QSOs (see Table \ref{tab:QSOinfo}). 
Additionally, we find that all parameters but the Eddington ratio ($L_\text{Bol}$, SFR, IR luminosity, dust mass) increase with $M_\text{BH}$ from \ourq\  through J0305$-$3150 and J1030$+$0524 to J1148$+$5251.
If a larger $M_\text{BH}$ means a longer elapsed time, the increase in these physical parameters with $M_\text{BH}$ 
may
naively indicates their evolution. A bright \Lya halo may appear only in an early phase of black hole growth.

A black hole evolves with its host galaxy, that is so called co-evolution (e.g., \citealp{maroggian98,merritt01,maclu02,marconi03,graham13,kormen13} and reference therein). 
According to the theoretical framework of galaxy evolution (e.g., \citealp{hopkins08}), the accretion disk around a black hole becomes visible in the rest-UV to optical wavelengths (and observed as a QSO) when the dust surrounding it is destroyed or blown away by negative feedbacks. After that, the QSO actively radiates UV emission with a high Eddington ratio. 
We call the phase when the QSO has just started to shine in UV the $young\ QSO\ phase$. 
Then the QSO is supposed to evolve by increasing its stellar mass and dust mass, as well as black hole mass, by acquiring gas from the IGM ($mature\ QSO\ phase$). Finally, the QSO's activity will weaken because the gas is consumed and dispersed ($old\ QSO\ phase$).
In this scenario, an extended and bright \Lya halo exists only in the $young\ QSO\ phase$ where a large amount of ionizing photons from the QSO easily escape to the circum-galactic region without strong dust extinction, and ionize neutral hydrogen of the CGM. In the $mature$ and $old\ phases$, on the other hand, an enormous amount of dust in the inter-stellar medium (ISM) makes ionizing photons difficult to leak away, thus preventing the CGM, even if it exists, from radiating \Lya emission. Thus, both the presence of a \Lya halo and the trends seen in the physical parameters of the four $z>6$ QSOs can be consistently interpreted in the evolutionary framework of QSOs. 
\ourq\  with a distinguishing \Lya halo owing to a small amount of dust is probably in the $young\ QSO\ phase$, while the other three with massive black holes ($M_\text{BH} > 10^9$ M$_\odot$) and high dust masses ($M_{dust} > 10^8$ M$_\odot$) are likely in the $mature$ or $old\ QSO\ phase$.

Although we find a negative trend that the luminosity and size of \Lya halos decrease with dustiness and $M_\text{BH}$ among four $z>6$ QSOs, this may happen by chance. In order to confirm the presence of this trend, observations of \Lya emission targeting $z>6$ QSOs with dustiness and $M_\text{BH}$ estimates are needed.

\subsubsection{Physical properties of QSOs and their \Lya halos at low-$z$}

We discuss the relation between \Lya halo properties and QSO properties at $z\sim2-3$.
In Section \ref{sec:QSO_dustiness}, we find that there are possible weak tendencies especially between \Lya halo scales ($d$ and $L_{Ly\alpha}$/$L_{Bol}$) and the IRX, although Spearman's $\rho$ values have large error bars due to a small sample size. This trend is comparable with the one seen for the $z>6$ QSOs over a wider IRX range. It may imply that dust abundance is also key to producing a \Lya halo around low-$z$ QSOs.
However, again, this possible negative trend may be statistically insignificant because it is based on only several QSOs. Thus we need more measurements of both IRX and \Lya halo scales to low-$z$ QSOs. Fortunately, \Lya halos have already been detected for dozens of QSOs at $z\sim3$ in the literature \citep{borisova16b,arrigoni18b}. 
IR or radio data of these QSOs would provide IRX values, and evaluate the correlation between dustiness and \Lya halo scales.

We also find that some low-$z$ QSOs with massive black holes ($M_\text{BH}>10^9$ $M_\odot$) have relatively large, $d<100$ pkpc, \Lya halos (see Figure \ref{fig:QSO_host}), whereas none of the $z>6$ QSOs with the same mass range has such a halo, despite the fact that all the QSOs have similar trends in $d$ vs IRX and $L_{Ly\alpha}$/$L_\text{Bol}$ vs IRX.
Therefore, we infer that the discrepancy of the presence/absence of \Lya halos at $M_\text{BH}>10^9$ M$_\odot$ is due to differences in the evolutionary track, and/or differences in physical properties of the ISM, such as dust density, gas mass/density, and the size of the  star-forming disk.

In Sections 5.1.1 and above, we have found possible negative correlations of \Lya luminosity with $M_\text{BH}$ or IRX, and proposed a hypothesis that QSOs in younger phase have larger \Lya halos.
Note, however, that the following two possibilities cannot be ruled out.
First is that the observed correlations have arisen just by chance due to small samples. In particular, the $z>6$ sample consists of only four objects among which two have only an upper limit of $L_{Ly\alpha}$.
The second possibility is that the two negative correlations are real but they are caused by other physics which we do not consider, because dust abundance is not the only factor to determine the size and luminosity of an extended \Lya halo.
If dust has a clumpy distribution, ionizing photons can escape from the host galaxy more easily, making a large and bright \Lya halo.
It has also been suggested that a wider opening angle gives a high escape fraction of ionizing photons (e.g. \citealp{trainor13}; but see also \citealp{henna13}).
The resonant scattering of \Lya photons by  \ion{H}{i} gas (if any) in the CGM can also contribute to extended halos, where the escape fraction of \Lya photons is determined by the  \ion{H}{i} gas distribution and dynamics as well as the dust distribution in the inter-stellar medium (ISM) (e.g. \citealp{laursen07,laursen09b,dijk12,verha12,yajima18}). 
It may not be easy to reproduce the observed correlations with these physical factors, because one has to assume, for example, that opening angle is dependent on $M_\text{BH}$ and IRX. 
However, the current data cannot rule out such dependencies. We will leave further discussion of this issue for future work.

\subsection{The CGM around QSOs}

\subsubsection{Is the CGM of \ourq\  optically thin or thick?}
\label{thinj2329}

\citet{henna13} have proposed several models which can explain the observed SB and \Lya luminosity of \Lya halos around QSOs by assuming different origins of \Lya emission. 
Because QSOs' \Lya halos are known to be mainly due to fluorescence (e.g., \citealp{hogman87,rees88,gould96,haiman01,alam02,cantalupo05,koll10}), here we consider their fluorescence model. 
Here we examine the optical thickness of \ourq's halo by following the same approach as in \citet{farina17}.
First, we test the case that \ourq\ is surrounded by optically thick gas. In this scenario, \Lya emission produced through recombination is mainly from the skin of clouds, and can be quantified with
\begin{equation}
 	\frac{L_{Ly\alpha}}{10^{44} \text{ erg s$^{-1}$}} = 7.8 \times f_c^{\text{thick}} \frac{L_{\text{$\nu$LL}}}{10^{30} \text{ erg s$^{-1}$ Hz$^{-1}$}},
\end{equation}
where $L_{Ly\alpha}$ is the total \Lya luminosity of the halo, $f_c^{\text{thick}}$ the covering factor for optically thick clouds, and $L_{\text{$\nu$LL}}$ the ionizing luminosity evaluated at the Lyman limit \citep{henna13,farina17}.
Then, the $L_{Ly\alpha}$ above which the CGM is in the optically thick regime is derived as $L_{Ly\alpha}=3.1\times10^{44}$ erg s$^{-1}$ by using the ionizing luminosity\footnote{
The $L_{\text{$\nu$LL}}$ is derived using the relations found from QSOs stacked spectra in \citet{lusso15} with the $1450$ {\AA} absolute magnitude of \ourq\  $M_{1450} = -25.25$.
}
$L_{\text{$\nu$LL}}=4.0\times10^{30}$ erg s$^{-1}$ Hz$^{-1}$, and $f_c^{\text{thick}}=0.1$ as has been obtained for small-scale \Lya emission found around $z=2-3$ QSOs \citep{henna13}.
This $L_{Ly\alpha}$ value is about a factor of three larger than the observed value, $L_{Ly\alpha}=9.8\times10^{43}$ erg s$^{-1}$. 
Thus, in order for \ourq's \Lya halo to be optically thick, $f_c^{\text{thick}}$ must be relatively low, below the observed value of $0.03$.
However, optically thick gas around QSOs is reported to have higher $f_c^{\text{thick}}$ values of $f_c^{\text{thick}}>0.2$ by an independent method \citep{prochas13}. Therefore, optically thick gas is not probably the case. We should, however, note that optically thick gas with an unusually small covering fraction cannot be ruled out.

Next, we consider the case that the CGM is optically thin. 
If a QSO is surrounded by optically thin gas, it would be sufficiently ionized by ionizing photons from the QSO.
Gas of a \Lya halo is considered to be optically thin if the neutral column density averaged over the halo $\langle N_{\ion{H}{i}}\rangle$ is less than $10^{17.2}$ cm$^{-2}$ \citep{henna13}. 
We evaluate $\langle N_{\ion{H}{i}} \rangle$ of \ourq's halo. According to \citet{henna13}, $\langle N_{\ion{H}{i}} \rangle$ is expressed by
\begin{eqnarray}
	\label{eq:avecloumnden}
	\frac{\langle N_{\ion{H}{i}} \rangle}{10^{17.2} \text{cm}^{-2}} =  1.1\  \biggl( \frac{L_{Ly\alpha}}{10^{44}  \text{erg s$^{-1}$}} \biggl) \biggl(\frac{L_{\text{$\nu$LL}}}{10^{30} \text{ erg s$^{-1}$ Hz$^{-1}$}}\biggl)^{-1} .
\end{eqnarray}
We obtain $\langle N_{\ion{H}{i}} \rangle = 10^{16.6}$ cm$^{-2}$,
which is a factor of $3.7$ lower than the threshold value.
Again, \ourq\  requires a very small $f_c^{\text{thick}}$ value of $<0.03$ to regard its CGM as optically thick, that is inconsistent with $f_c^{\text{thick}}$ found around low-$z$ QSOs \citep{prochas13}.
Therefore, these two tests indicate that the optically thin CGM is probably the case for \ourq.

\subsubsection{Is the CGM of the other QSOs optically thin or thick?
}
\label{thin_compsample}
 
We also use Equation \ref{eq:avecloumnden} to evaluate  $\langle N_{\ion{H}{i}} \rangle$ for the other QSOs in the compilation sample.
Because not all objects have $L_{\text{$\nu$LL}}$ measurements, 
we evaluate $L_{\text{$\nu$LL}}$ by scaling the value of \ourq\ by the $L_\text{Bol}$ ratio between the object in question and \ourq\ assuming that 
$L_{\text{$\nu$LL}}$ is proportional to $L_\text{Bol}$.
Among the objects in the compilation sample, $47\%$ ($68$ QSOs) have both $L_\text{Bol}$ and $L_{Ly\alpha}$ measurements, and we find almost all of them to be optically thin.
For those without $L_\text{Bol}$ measurements but with  $L_{Ly\alpha}$ data ($44\%$ or $64$ objects of the compilation sample),
we just use the $L_{\text{$\nu$LL}}$ of \ourq, and again find that almost all are optically thin.
If an object is fainter in $L_{\text{$\nu$LL}}$ but brighter in $L_{Ly\alpha}$ than \ourq, 
its CGM would be regarded as optically thick. However, the fraction of such objects is likely to be less than $31\%$; $45$ objects are more luminous in $L_{Ly\alpha}$ than \ourq\ and only $2$ have higher $L_{Ly\alpha}$ and lower $L_\text{Bol}$ than \ourq. 
We cannot evaluate $\langle N_{\ion{H}{i}} \rangle$ for the remaining 
$8\%$
of the compilation sample which have neither $L_{Ly\alpha}$ or $L_\text{Bol}$.
To summarize, at least 
$88\%$
of the compilation sample are probably optically thin.

\begin{figure*}
	\begin{center}
	\includegraphics[width=\linewidth]{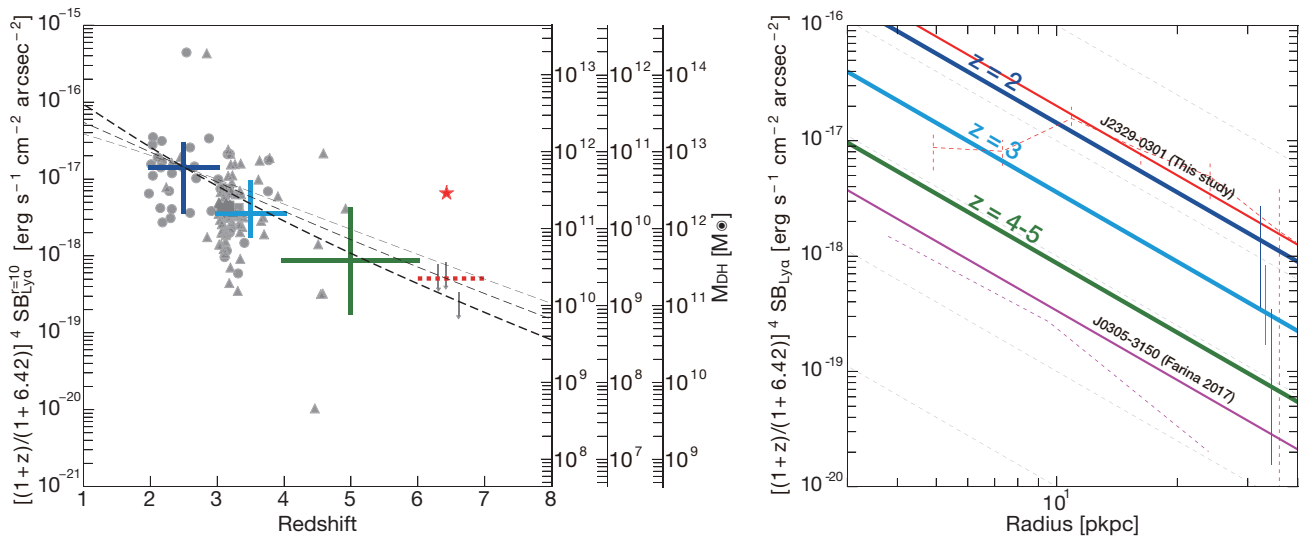} 
	\caption{(Left) SB$_{Ly\alpha}^{r=10}$ vs redshift. 
	Colored cross symbols represent median values with errors for $z=2$, $3$, and $4-5$ bins. A dashed red cross is an expected SB$_{Ly\alpha}^{r=10}$. 
	Three dashed lines represent SB$_{Ly\alpha}^{r=10}$ which scales with the mass of dark halos growing up to $1\times10^{11}$ (thin), $1\times10^{12}$ (middle) and $1\times10^{13}$ M$_\odot$ (thick) at $z=2$.
	The ordinate in the right hand side is the dark halo mass calibrated so that $1 \times 10^{12}$ M$_\odot$ corresponds to the median SB$_{Ly\alpha}^{r=10}$ at $z=2$.
	(Right) Same as Figure \ref{fig:radip}, but for median SB profiles derived from median SB$_{Ly\alpha}^{r=10}$ values on the assumption of SB$(r) \propto r^{-2}$: blue, $z=2-3$; cyan, $z=3$; green, $z=4-5$. Vertical lines colored in blue, cyan and green are the $1\sigma$ of the data distributions.
	Red and purple thin straight lines represent the SB profiles of \ourq\  and J0305$-$3150 derived from SB$_{Ly\alpha}^{r=4}$ and SB$_{Ly\alpha}^{r=2}$, respectively, while dashed lines are their original profiles.  
	(A color figure is available in the online journal.)
	}
	\label{fig:SB10}
	\end{center}
\end{figure*}


\subsubsection{Possible redshift evolution of the CGM}
\label{sec:CGM_QSO}

We find that $L_{Ly\alpha}$, $L_{Ly\alpha}$/$L_\text{Bol}$ and $d$ of \Lya halos decrease with redshift. As described in Section \ref{comp_z}, these decreasing trends seem to be real because no QSO at $z>4$ has an extended \Lya halo despite having similarly bright $L_\text{Bol}$ to lower-$z$ objects.
Furthermore, the amplitude of the SB profile appears to decrease with redshift as shown in Figure \ref{fig:radip} (b) except that of \ourq\  \citep{borisova16b,farina17}.
The redshift evolution of $L_{Ly\alpha}$, spatial extent and SB amplitude has also been discussed in the literature \citep{north12,farina17,ginolfi18}. 
These pieces of evidence together with our finding that there is no bright and extended \Lya halos at $z>4$ appear to indicate some evolution of the CGM.

In order to obtain further insights, we compare the characteristic SB profile of \Lya halos among three redshift bins of $2<z<3$, $3<z<4$, and $4<z<6$ that denote $z=2$, $3$, and $4-5$. The SB profiles of $z>6$ QSOs are discussed separately because of a very small sample size. Since not all objects in the compilation sample have SB data, we derive the characteristic SB profile at each redshift bin in the following manner.
\begin{itemize}
	\item[i)] For each object, we calculate the SB at $10$ pkpc radius (SB$_{Ly\alpha}^{r=10}$) from $L_{Ly\alpha}$ and $d_\text{org}$. We choose $10$ pkpc radius, since it is large enough to be within halos, being close to their inner most radii (e.g., \citealp{borisova16b}), and small enough to have high-$S/N$ \Lya emission. 
	Because of this definition for deriving SB$_{Ly\alpha}^{r=10}$, we only use objects with $d_\text{org} \ge 20$ pkpc (see also Table \ref{tab:sampleinfo}). SB$_{Ly\alpha}^{r=10}$ values after correction for SB dimming are shown in Figure \ref{fig:SB10} (left). \\
	\item[ii)] We then use SB$_{Ly\alpha}^{r=10}$ values to calculate the median SB$_{Ly\alpha}^{r=10}$ for each redshift bin (blue, cyan and green crosses in Figure \ref{fig:SB10} left). Because SB$_{Ly\alpha}^{r=10}$ values are widely distributed from $10^{-21}$ to $10^{-15}$ erg s$^{-1}$ cm$^{-2}$ arcsec$^{-2}$, we adopt not an average but a median for the characteristic SB, which is found to be
	$\langle$SB$_{Ly\alpha}^{r=10} \rangle =$ ($1.4\times10^{-17}$, $3.6\times10^{-18}$, $8.7\times10^{-19}$) erg s$^{-1}$ cm$^{-2}$ arcsec$^{-2}$ for $z=(2$, $3$, $4-5$).\\
	\item[iii)] Finally, we obtain the characteristic SB profile at each redshift bin from the characteristic SB$_{Ly\alpha}^{r=10}$ assuming a power-law profile of SB $\propto r^{-2}$ to simplify the calculation.
	We also confirm that the average SB profile obtained by our method is consistent with the one presented in \citet{borisova16b}. 
	The SB profiles corrected for cosmological dimming are shown as thick blue, cyan and green lines in Figure \ref{fig:SB10} (right). \\
	\item[iv)] For two QSOs at $z>6$ with a detected \Lya halo, \ourq\  and J0305$-$3150, we estimate SB$_{Ly\alpha}^{r=10}$ by fitting a power law to the data points where \Lya emission is significantly detected, because we find that SB profiles calibrated at SB$_{Ly\alpha}^{r=10}$ slightly overestimate the true profiles. The profiles thus obtained are plotted as thin red and purple lines in Figure \ref{fig:SB10} (right). 
	The SB$_{Ly\alpha}^{r=10}$ values are ($6.5\times10^{-18}$, $3.4\times10^{-19}$) erg s$^{-1}$ cm$^{-2}$ arcsec$^{-2}$ for (\ourq, J0305$-$3150).
	For the remaining two $z>6$ QSOs whose \Lya halos are undetected, we only estimate an upper limit.
\end{itemize}

We find that the characteristic SB profile clearly decreases with redshift (Figure \ref{fig:SB10} right), although the sample sizes are still small at $z>4$. The amplitude of the characteristic SB profile at $z=2$ is about a factor of $4.0$ ($16.6$) brighter than that at $z=3$ ($z=4-5$). 
We also compare the SB profiles of the two $z>6$ halos with those at lower redshifts. 
Surprisingly, \ourq's SB is as bright as that of $z=2$ halos, while J0305$-$3150's SB is even fainter than those at $z=4-5$.

In the optically thin regime of \citet{henna13} model 
which appears to be valid for most of our objects, the \Lya SB is proportional to the hydrogen volume density $n_\text{H}$, the hydrogen total column density $N_\text{H}$, and the covering factor of clouds in the optically thin CGM $f_C^\text{thin}$ (i.e., SB$_{Ly\alpha}\propto n_\text{H} N_\text{H} f_C^\text{thin}$). 
Thus, a decrease in the SB amplitude with redshift implies a decrease in the ``hydrogen density'', $n_\text{H} N_\text{H}$, or $f_C^\text{thin}$, or both.
In addition, in the optically thin regime, the total gas mass of the CGM is proportional to $N_\text{H}$ and $f_C^\text{thin}$. 
Hence, the decrease in the SB amplitude likely implies a decrease in the total gas mass of the CGM.

\subsubsection{The relation between \Lya halo scales and the dark matter halos of hosting QSOs}
\label{sec:M_DM}

If the CGM mass around QSOs increases with cosmic time as suggested in Section \ref{sec:CGM_QSO}, what physics causes that?
One possibility is the growth of hosting dark matter halos.

Cosmological simulations predict that QSOs at higher-$z$ reside in less massive dark matter halos (e.g., \citealp{fani12,oogi16}). This prediction is also supported by observations (e.g., \citealp{eft15,he18,uchi18}).
Here we use a fitting formula given in \citet{behrooz13} to estimate the mass growth rate of QSO hosting halos at $z>2$, assuming that they have grown to $1\times10^{11}$, $1\times10^{12}$ or $1\times10^{13}$ M$_\odot$ at $z=2$, among which $1\times10^{12}$ M$_\odot$ is the closest to the observed values (e.g., \citealp{adel05b,white12}). 
Surprisingly, these halo growth rates, especially those of $1\times10^{12}$ and $1\times10^{13}$ M$_\odot$, are in good agreement with those of SB$_{Ly\alpha}^{r=10}$, or the amplitude of the characteristic SB profile:
the SB at $z=2$ is factors $4.0$ and $16.6$ higher than those at $z=3$ and $z=4-5$, respectively. 
In other words, the amplitude of the SB scales with the mass of evolving dark matter halos (dashed lines in Figure \ref{fig:SB10} left).
Thus, {\it if} the QSOs at different redshifts are on average in a progenitor-descendant relationship in terms of hosting dark matter halos and {\it if} the SB amplitude is proportional to the total mass of the CGM, then the coincidence found here implies that the CGM grows in mass keeping pace with hosting dark matter halos.

We then examine the evolution of the extent of \Lya halos using characteristic SB profiles.
We define the extent as the diameter of \Lya halos at a level of [($1+z$)/($1+6.42$)]$^4$ SB$_{Ly\alpha} = 10^{-18}$ erg s$^{-1}$ cm$^{-2}$ arcsec$^{-2}$, and obtain 
($76$, $38$, $20$) pkpc for $z=$($2$, $3$, $4-5$).
In Figure \ref{fig:size}, we compare these characteristic extents with two scaling models.
One is ($1+z$)$^{-1}$ scaling, i.e., scaling with the virial radius of dark halos with a constant mass (dotted line in Figure \ref{fig:size}) as proposed by \citet{ginolfi18}. 
The other is $M_\text{DH}^{1/3}$($1+z$)$^{-1}$ scaling, i.e., scaling with the virial radius of evolving dark halos which have 
$1 \times 10^{11}$, $1 \times 10^{12}$ or $1 \times 10^{13}$ M$_\odot$ at $z=2$.
All models have been calibrated at $z=2$.
We find that $M_\text{DH}^{1/3}$($1+z$)$^{-1}$ scaling agrees well with the data, while ($1+z$)$^{-1}$ scaling is not steep enough.
We stress that the point of our findings here is the assumption that the QSOs in the compilation sample have a progenitor-descendant relationship in terms of hosting dark halos. As shown in Figures \ref{fig:SB10} (left) and \ref{fig:size}, our findings are not sensitive to changes in $M_\text{DH}(z=2)$.
It is very interesting that the evolution of the SB and extent of \Lya halos can be simultaneously explained by a simple scenario that QSOs are on average in a progenitor-descendant relationship in terms of hosting dark halos and that the mass and size of the CGM just scale with those of hosting halos.

Finally, by extrapolating these evolutionary trends found over $z \sim 2-5$, we predict the $\langle$SB$_{Ly\alpha}^{r=10} \rangle$ and the characteristic extent for $z\sim6$ \Lya halos hosted by progenitors of $z=2$ dark halos with $1\times10^{12}$ M$_\odot$. Since the dark halo mass growth rate from $z=6$ to $z=2$ is calculated to be $31.3$ using the formula of \citet{behrooz13}, we obtain $\langle$SB$_{Ly\alpha}^{r=10} \rangle = 4.6\times10^{-19}$ erg s$^{-1}$ cm$^{-2}$ arcsec$^{-2}$ and the characteristic extent as $14$ pkpc as shown as dotted red symbols in Figures \ref{fig:SB10} (left) and \ref{fig:size}.
These values are comparable to those of J0305$-$3150, but much lower than those of \ourq. This may suggest that \ourq\ is a rare QSO which has an exceptionally bright \Lya halo, although a statistically meaningful comparison requires a much larger sample.

%
\subsubsection{Systematic uncertainties and biases that could cause a redshift evolution of QSOs' \Lya halos}

In this subsection, we discuss systematic uncertainties and biases that could produce (part of) the observed decline of SB$_{Ly\alpha}^{r=10}$ (Section \ref{sec:CGM_QSO}).
Four candidates are discussed.

The first one is underestimation of $L_{Ly\alpha}$. 
The \Lya fluxes of halos in our compilation sample have been measured by either slit spectroscopy or imaging (including integral field unit or integral field spectroscopy); those from slit spectroscopy are fainter than total \Lya fluxes because of slit loss.
Although the original papers have given slit-loss corrected \Lya fluxes, those corrections may be insufficient.
For instance, we find that \ourq's $L_{Ly\alpha}$ derived from spectroscopic observations in \citet{willo11} is smaller than our estimate by a factor of $1.2$. If such underestimation is present in our compilation sample, the $true$ characteristic median of $\langle$SB$_{Ly\alpha}^{r=10}\rangle$ should be larger than our estimates. In particular, the $L_{Ly\alpha}$ of most of the $z=4-5$ halos have been measured by slit spectroscopy. 
Thus, we evaluate the sensitiveness of $\langle$SB$_{Ly\alpha}^{r=10}\rangle$ to the slit-loss correction factor of those $z=4-5$ halos.
We find that the $\langle$SB$_{Ly\alpha}^{r=10}\rangle$ at $z=4-5$ becomes as bright as the $z=3$ value if their $true$ $L_{Ly\alpha}$ were larger by about a factor of $10$.
Such a large amount of slit loss appears to be unrealistic.

The second candidate is a possible lack of ELAN-like halos at $z=4-5$. At $z=2-3$, extremely bright and extended \Lya halos have been found around some QSOs, and are called ELAN (Enourmous Lyman-Alpha Nebulae; \citealp{cantalupo14,henna15,cai17,Cai_2018,arrigoni18b,arrigoni18}). Current statistics of the detection probability of ELAN suggest its rareness of only a few percent of QSOs at $z\sim3$ \citep{arrigoni18b}. 
If ELAN were similarly rare at $z=4-5$ and our compilation sample misses such ELAN-like halos, our $\langle$SB$_{Ly\alpha}^{r=10}\rangle$ value would be biased low.
To examine the effect of ELAN on $\langle$SB$_{Ly\alpha}^{r=10}\rangle$ estimates, we exclude ELAN from our compilation sample, and derive $\langle$SB$_{Ly\alpha}^{r=10}\rangle$. The $\langle$SB$_{Ly\alpha}^{r=10}\rangle$ values thus obtained are ($1.4\times10^{-17}$, $3.5\times10^{-18}$, $8.7\times10^{-19}$) erg s$^{-1}$ cm$^{-2}$ arcsec$^{-2}$ for $z=(2$, $3$, $4-5$); the differences from the original values are very small. 
Thus, the presence of ELAN has negligible effects on our results.

Our SB and size estimates will also be biased low if our sample misses relatively large \Lya halos with $d>100$ pkpc at $z>4$.
Indeed, radio-quiet QSOs at $z\sim2$ are known to possess no or only a small halo with $d<50$ pkpc (e.g. \citealp{herenz15,arrigoni16}), and all the $z=4-5$ objects in our sample are radio-quiet (see also Figure \ref{fig:SB10} left). However, currently there is no observation that suggests $z=4-5$ QSOs to have similar radio-loudness dependence.
A larger sample including radio-loud QSOs is needed for further discussion.

The third candidate is a small sample size at $z=4-5$. Our $\langle$SB$_{Ly\alpha}^{r=10}\rangle$ value at this redshift bin is derived from only six QSOs, which are much fewer than those at $z=2$ and $3$. 
Our $z=4-5$ sample may be greatly underestimating the true $\langle$SB$_{Ly\alpha}^{r=10}\rangle$ owing to large statistical fluctuations.
However, we cannot test this possibility by our sample.
A larger sample is needed to do so.

Finally, we discuss the possibility that the decreasing trend of $\langle$SB$_{Ly\alpha}^{r=10}\rangle$ is not a common feature of all QSOs but seen only in those satisfying our selection criteria.
Because the purpose of this study is to investigate correlations between \Lya halo properties and hosting QSOs' physical parameters, we do not include QSOs whose halo is either undetected or too small (diameter $<20$ pkc) to define SB$_{Ly\alpha}^{r=10}$(see also Section \ref{sec:CGM_QSO}). In this respect, our results are biased.
However, \citet{arrigoni18b} have found an opposite trend that the SB at $z=2$ is a factor of $10$ fainter than that at $z=3$ when they have included in the $z=2$ sample those with an undetected or small \Lya halos.
In fact, not all $z=2$ QSOs have \Lya halos, with detection rates of about $30-70\%$ (e.g., \citealp{leh99,heckman91b,henna13,arrigoni16}). 
Including such QSOs in our sample should naturally reduce the median SB.
However, it is not straightforward to estimate the degree of the reduction, because \citet{arrigoni18b} have not included large detected \Lya halos at $z=2$. In addition, non-detected \Lya halos may also be present at $z=4-5$ and may similarly reduce the median SB there.
Tracing the \Lya halo evolution of the entire QSO population needs an unbiased survey like the QSO MUSEUM (Quasar Snapshot Observations with MUse:Search for Ex-tended Ultraviolet eMission: \citealp{arrigoni18b}) over a wide redshift range.

\begin{figure}
	\begin{center}
	\includegraphics[width=\linewidth]{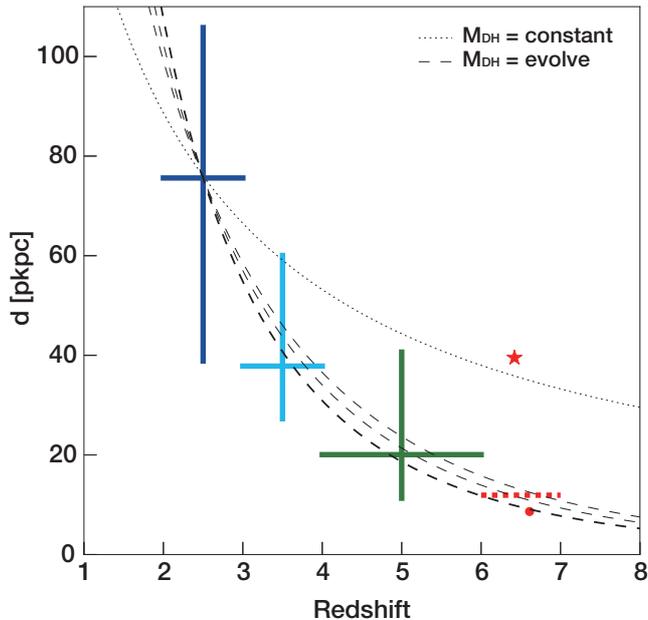} 
	\caption{Extent of \Lya halos as a function redshift. Colored cross symbols represent the characteristic extents at $z=2,\ 3,\ 4-5$ defined as the diameter of halos at [($1+z$)/($1+6.42$)]$^4$ SB$_{Ly\alpha} = 10^{-18}$ erg s$^{-1}$ cm$^{-2}$ arcsec$^{-2}$. Because the characteristic extent at $z=6$ is a prediction, it is plotted with a dotted line. A star and a circle represent the extents of \ourq\ and J0305$-$3150, respectively.
	A dotted line shows scaling with ($1+z$)$^{-1}$, while three dashed lines are scaling with $M_\text{DH}^{1/3}$($1+z$)$^{-1}$ as in the case of Figure \ref{fig:SB10} (left).
	(A color figure is available in the online journal.)
	}
	\label{fig:size}
	\end{center}
\end{figure}


\section{Summary}
\label{summary}

In this paper, we have first investigated the very luminous \Lya halo around the QSO \ourq\ at $z=6.42$ with new data. Then, we have systematically studied the properties of QSO \Lya halos over $z \sim 2-6$ using all available data in the literature (`compilation sample'). The major results are summarized below.

\begin{enumerate}
	\renewcommand{\theenumi}{\arabic{enumi}.}
	\item We have confirmed extended \Lya emission around \ourq, with a \Lya luminosity of 
	$9.8\times10^{43}$ erg s$^{-1}$ within an extent of $\sim37$ pkpc.
	The SB of this halo is about an order of magnitude brighter than that of another halo-detected QSO at $z>6$, QSO J0305$-$3150, but comparable to those of luminous halos at $z\sim3$ \citep{borisova16b}. 
	\item We have examined correlations of several parameters characterizing \Lya halos with [1] redshift, [2] $M_\text{BH}$ the Eddington ratio, and $L_\text{Bol}$ and [3] spectral index $\alpha$ and IRX among the compilation sample. 
	\begin{enumerate}
		\item[[ 1]] We have found declines of $L_{Ly\alpha}$, $L_{Ly\alpha}$/$L_\text{Bol}$ and $d$ with redshift, indicating some redshift evolution of \Lya halos. 
		\item[[ 2]] We have found a strong positive correlation between $L_{Ly\alpha}$ vs. $d$ and a moderate correlation between $d$ vs. $L_{Ly\alpha} / L_\text{Bol}$ based on Spearman's test.
		A possible negative trend has also been seen in the $d$ vs. $M_\text{BH}$ and $d$ vs. $L_{Ly\alpha} / L_\text{Bol}$ distributions when limited to $z>6$.
		\item[[ 3]] Spearman's test has also indicated a moderate negative correlation of $d$ and $L_{Ly\alpha}$/$L_\text{Bol}$ with the IRX, although the errors in the $\rho$ parameter are large due to a very small sample size. 
		These trends become relatively clearer when the sample is limited to $z>6$. 
	\end{enumerate}
	\item We have examined physical properties of four $z>6$ QSOs which have a wide range of $L_{Ly\alpha}$.
	We have found that the dust content probably controls the presence/absence of \Lya halos because QSOs with no or a small \Lya halo have a higher IR luminosity and dust mass than \ourq\ which has an extraordinarily luminous and extended \Lya halo. 
	\ourq\ also has the least massive $M_\text{BH}$ ($M_\text{BH} < 10^9$ M$_\odot$) among the four. 
    We infer that QSOs have a \Lya halo only in the young phase because those in older phases have a large amount of dust which absorbs ionizing photons before escaping out to circum-galactic regions. 
	\item The \Lya halo around \ourq\ is optically thin against ionizing photons. It is also found that at least $\sim 88\%$ of the compilation sample are optically thin.
	\item We have derived the characteristic SB profile at $z=2$, $3$, and $4-5$ from the SB at $10$ pkpc radius (SB$_{Ly\alpha}^{r=10}$) with an assumption of a universal power-law profile of SB $\propto r^{-2}$, and then, found that SB$_{Ly\alpha}^{r=10}$ increases with cosmic time. Its growth rate between $z=2$ and $z=3$ ($z=4-5$) is a factor of $4.0$ ($16.6$).
	Because the SB is proportional to ``hydrogen density ($n_\text{H} N_\text{H}$)'' and $f_C^\text{thin}$, and the total gas mass of the CGM scales with these parameters, the brightening of the SB likely indicates an increase in the CGM gas mass with time.
    The increasing rate of the SB coincides with the mass growth rate of dark halos that evolve to $1 \times 10^{12}$ or $1\times10^{13}$ M$_\odot$ at $z=2$. 
	\item We have also estimated the characteristic extents of \Lya halos to be 
	($76$, $38$, $20$)
	pkpc for $z=$($2$, $3$, $4-5$). The evolution of the characteristic extent does not match ($1+z$)$^{-1}$ scaling which is suggested by \citet{ginolfi18}, but matches well $M_\text{DH}^{1/3}$($1+z$)$^{-1}$ scaling, i.e., scaling with the virial radius of evolving dark halos.
	These increases in SB and extent with time are consistent with a scenario that the CGM around QSOs evolves in mass and size keeping pace with hosting dark matter halos. 
	\item Extrapolating these evolutionary trends, we have predicted the mean SB$_{Ly\alpha}^{r=10}$ and extent for $z=6$ \Lya halos to be  $\langle$SB$_{Ly\alpha}^{r=10} \rangle = 4.6\times10^{-19}$ erg s$^{-1}$ cm$^{-2}$ arcsec$^{-2}$ and $14$ pkpc. These values are comparable to those of J0305$-$3150, but much lower than those of \ourq, indicating a rareness of \ourq's \Lya halo.
\end{enumerate}

\section*{Acknowledgements}
We appreciate the referee and the editor for providing the constructive suggestions and comments to improve our manuscript. 
We are grateful to E. P. Farina and F. Arrigoni Battaia for providing the data of \citet{farina17} and \citet{arrigoni18b}, and S. Cantalupo, G. Pezzulli and E. Borisova for providing the data of \citet{borisova16b}.
We thank M. Rauch, Z. Cai, J. X. Prochaska, K. Inayoshi, M. Onoue, T. Oogi, H. Kusakabe, T. Okamura, S. Mukae, and M. Ando for helpful discussions.
We acknowledge grant aid for the narrow band filter from the Department of Astronomical Sciences of the Graduate University for Advanced Studies (SOKENDAI). 
RM acknowledges a Japan Society for the Promotion of Science (JSPS) Fellowship at Japan and a Ministry of Science and Technology (MOST) Fellowship at Taiwan. RM was supported by JSPS KAKENHI 18J40088, and by MOST grant 104-2112-M-007-021-MY3. 
YU was supported by JSPS KAKENHI Grant Number JP26800103, JP24103003.
TG acknowledges support by the Ministry of Science and Technology of Taiwan through grant 105-2112-M-007-003-MY3.


\bibliographystyle{mnras}
\bibliography{J2329}

\end{document}